\documentclass{PoS}

\title{When epsilon-expansion of hypergeometric functions is expressible in terms of multiple polylogarithms: the two-variables examples}
\ShortTitle{The structures of higher hypergeometric functions and the epsilon-expansion}

\author{Vladimir~V.~Bytev\\
Joint Institute for Nuclear Research,$141980$ Dubna (Moscow Region), Russia \\
E-mail: \email{bvv@mail.jinr.ru}}

\author{\speaker{Mikhail Kalmykov}%
        \thanks{
This work was supported in part by the German Federal Ministry for Education
and Research BMBF through Grant No.\ 05~HT6GUA, by the German Research
Foundation DFG through the Collaborative Research Centre No.~676
{\it Particles, Strings and the Early Universe---The Structure of Matter and
Space-Time}, and by the Helmholtz Association HGF through the Helmholtz
Alliance Ha~101 {\it Physics at the Terascale}.
}\\
II. Institut f\"ur Theoretische Physik, Universit\"at Hamburg,
Luruper Chaussee 149, 22761 Hamburg, Germany \\
Joint Institute for Nuclear Research,$141980$ Dubna (Moscow Region), Russia \\
E-mail: \email{kalmykov.mikhail@gmail.com}}

\author{Bernd~A.~Kniehl \\
II. Institut f\"ur Theoretische Physik, Universit\"at Hamburg,
Luruper Chaussee 149, 22761 Hamburg, Germany \\
E-mail: \email{kniehl@mail.desy.de}}

\abstract{
In this talk, we discuss the algorithm for the construction of analytical 
coefficients of higher order epsilon expansion of 
some Horn type hypergeometric functions of two variables
around rational values of parameters.  
}

\FullConference{Loops and Legs in Quantum Field Theory - 11th DESY Workshop on Elementary Particle Physics \\
                 April 15-20, 2012\\
                 Wernigerode, Germany}

\begin{document}

\noindent
{\bf One-loop Feynman Diagrams and Hypergeometric functions}. \\
Within dimensional regularization \cite{dimreg}, 
the algorithm for analytical evaluation of one-loop multilegs Feynman Diagrams 
has been described a long time ago \cite{PV,tHV}. It includes two steps: reduction of 
the amplitude to a set of master-integrals \cite{PV} with following analytical evaluation of 
master-integrals via Feynman parameter representation \cite{tHV}. 
Due to the appearance of $1/\varepsilon$ terms coming from IR and/or UV singularities, 
the NNLO calculation would demand the knowledge of higher terms in the $\varepsilon$-expansion
of the one-loop Feynman Diagrams (see also \cite{Weinzierl}).
The above mentioned technology is suitable for the evaluation of the 
finite part of master-integrals \cite{1-loop,hexagon}.
However, the direct application of this technique for analytical evaluation of the 
higher-order coefficients in power of $\varepsilon$ gives rise to complicated results
\cite{one-loop:linear:1,one-loop:linear:2,pentagon} even in a simple kinematic.
A perspective approach to the construction of analytical coefficients of 
the $\varepsilon$-expansion of one-loop Feynman Diagrams 
is to explore the hypergeometric representation \cite{FJT,DD}.
Based on the approach developed in \cite{DD}, the all-order $\varepsilon$-expansion 
of one-loop self-energy diagrams has been constructed in $d=4-2\varepsilon$.
More examples of hypergeometric representation for one-loop diagrams based on the technique of \cite{FJT}
are given in \cite{vertex-ep,KT}.
However, still by now a systematic way to construct the analytical 
coefficients of the $\varepsilon$-expansion for Horn-type hypergeometric functions
around rational values of parameters does not exist.  

\noindent
{\bf Definition of the hypergeometric function}. \\
We remind the definition of {\it Horn-type Hypergeometric Functions}: \\
it is a formal (Laurent) power series in $r$ variables of the following form,
\begin{equation}
H(\vec{J};\vec{z}) 
\equiv 
H(\vec{\gamma};\vec{\sigma};\vec{z}) 
= 
\sum_{m_1,m_2,\cdots, m_r=0}^\infty 
\left(
\frac{
\Pi_{j=1}^K
\Gamma\left( \sum_{a=1}^r \mu_{ja}m_a+\gamma_j \right)
}
{
\Pi_{k=1}^L
\Gamma\left( \sum_{b=1}^r \nu_{kb}m_b+\sigma_k \right)
}
\right)
x_1^{m_1} \cdots x_r^{m_r} \;,
\label{Phi}
\end{equation}
with
$
\mu_{ab}, \nu_{ab} \in \mathbb{Q},\
\gamma_j,\sigma_k \in \mathbb{C}
$
and 
$\vec{J} \equiv \{\vec{\gamma_j}, \vec{\sigma} \}$.

%
%
%
\noindent
{\bf The problem under consideration:} \\ 
In arbitrary d-dimensional space time, 
$d=4-2\varepsilon$, 
where $\varepsilon$ is the parameter of dimension regularization \cite{dimreg}, 
the set of discrete parameters $\vec{J} \equiv \{\vec{\gamma_j}, \vec{\sigma} \}$ 
of hypergeometric function, 
$H(\vec{\gamma};\vec{\sigma};\vec{z})$
is a linear combination of rational and $\varepsilon$-dependent coefficients: 
$
J_k = A_{0,k} + a_k \varepsilon, 
$
where $A_{0,k}$ and $a_k$ are arbitrary rational numbers. 
The Laurent expansion of the hypergeometric function around the integer value $d=4$, 
is called 
``construction of the analytical coefficients of $\varepsilon$-expansion''
of the function:
\begin{equation}
H(\vec{A}_0 + \vec{a} \varepsilon; \vec{z})
=  
H(\vec{A}_0; \vec{z})
+ 
\sum_{j=1}^\infty \varepsilon^j h_j(\vec{z}) \;,
\label{expansion}
\end{equation}
where symbolically, 
\begin{eqnarray}
h_j(\vec{z}) = \left. \frac{\partial}{\partial \vec{A}} H(\vec{A}; \vec{z}) \right|_{\vec{A} = \vec{A_0}} \;.
\label{def:h}
\end{eqnarray}
The goal is to write the {\it coefficient functions} $h_j$ in terms 
of known functions, suitable for numerical evaluation \cite{VW}, 
or to describe all analytical properties of $h_j$, treating them as a new class of functions.

\noindent 
{\bf Existing algorithms}: \\
The first systematic algorithms for the construction of higher order coefficients 
of the $\varepsilon$-expansion of multivariable hypergeometric functions 
around integer values of parameters were suggested in \cite{MUW}. 
In Ref.~\cite{W}, the special set of rational values of parameters, 
the so called ``zero-balance case'' was analyzed.
However, the partial results of \cite{W} beyond the zero-balance case
are in contradiction with partial results of Ref.~\cite{DK02}.

\noindent 
{\bf Our method}: \\
In a series of papers \cite{our,KK2010,recent} it was shown 
that for the hypergeometric function of one variable, ${}_pF_{p-1}$,
the analytical coefficients of the $\varepsilon$-expansion can be constructed via 
an {\it explicit solution of differential equations} for coefficients functions $h_j(z)$.
Using Eq.~(\ref{def:h}) it is easy to show that the coefficients $h_j(\vec{z})$ 
satisfy the following linear system of (partial) differential equations (PDE):
\begin{eqnarray}
&& 
\sum_{\vec{L}} P_{\vec{L}} \frac{\partial^{\vec{L}}}{\partial \vec{z}} H(\vec{A};\vec{z}) = 0 
\Rightarrow
\left.  \frac{\partial}{\partial \vec{A}} 
\left[ 
\sum_{\vec{L}} P_{\vec{L}} \frac{\partial^{\vec{L}}}{\partial \vec{z}} H(\vec{A};\vec{z}) = 0 
\right]
\right|_{\vec{A} = \vec{A_0}} = 0 
\nonumber \\ && 
\Rightarrow
\left.  
\left[ 
\sum_{\vec{L}} P_{\vec{L}} 
\right]
\right|_{\vec{A} = \vec{A_0}}
\frac{\partial^{\vec{L}}}{\partial \vec{z}} h(\vec{z}) 
=  
- 
\left.  
\left[ 
\frac{\partial}{\partial \vec{A}} 
\sum_{\vec{L}} P_{\vec{L}} 
\right]
\right|_{\vec{A} = \vec{A_0}}
\left. 
\frac{\partial^{\vec{L}}}{\partial \vec{z}} H(\vec{A};\vec{z})  
\right|_{\vec{A} = \vec{A_0}} \;.
\label{h:de}
\end{eqnarray}
%
%
%
{\bf When are non-homogeneous PDE solvable in terms of multiple polylogarithms?} \\
We are interested in the question under which conditions the functions $h_j(z)$, solutions of Eq.~(\ref{h:de}),
are expressible in terms of {\it multiple polylogarithms} \cite{goncharov,1dm,2dm,abs}, or 
{\it generalized iterated integrals}, defined as:
\begin{eqnarray}
\hspace{-5mm}
G(z;R_k, R_{k-1}, \cdots , R_1) 
 =   
\int_0^{z} \frac{dt}{R_k(t)} I(t;R_{k-1}, \cdots, R_1)
 =   
\int_0^{z} 
\frac{dt_k}{R_k(t)} 
\int_0^{t_k} 
\frac{dt_{k-1}}{R_{k-1}(t)} 
\cdots 
\int_0^{t_2} \frac{dt_1}{R_1(t_1)} \;,
\label{gmp}
\end{eqnarray}
where 
$R_k(t)$ are some rational functions.
{\it Multiple polylogarithms} correspond to $R_k(t) = t-a_k$.
%
When is the system of PDE with non-zero non-homogeneous part solvable in terms of 
(generalized) multiple polylogarithms? {\bf Our algorithm} includes the following steps:
\begin{itemize}
\item
{\it Factorization}:
the differential operator(s) after $\varepsilon$-expansion 
are factorisable into product of differential operators of the first order; 
\item
{\it Linear parametrization} to all orders in $\varepsilon$; 
\item
The non-homogeneous part belongs to the class of functions of the special type (see below).
\end{itemize}

\noindent
{\bf Example I}\\
Let us consider the differential equation 
related to the hypergeometric function ${}_pF_{p-1}$ \cite{KK2010,recent}:
\begin{equation}
\sum_{k=0}^p P_k(z;\varepsilon) \left( \frac{d}{dz} \right)^k H(z;\varepsilon) = F(z;\varepsilon) \;, 
\label{de:example}
\end{equation}
where 
$P_k(z;\varepsilon)$ 
and 
$F(z;\varepsilon)$ 
are rational functions or iterated integrals over a rational 1-form:\\
\begin{equation}
P_k(z;\varepsilon) = \frac{\Pi_j(z-\alpha_j -\beta_j \varepsilon)}{\Pi_{r}(z-A_r- B_r \varepsilon)} \;,
\quad 
F(z;\varepsilon)  =  \int^z \frac{dt}{t-\sigma} \frac{\Pi_j(t-\mu_j -\nu_j \varepsilon)}{\Pi_{r}(t-M_r- N_r \varepsilon)} \;.
\end{equation}
We are looking for a solution of Eq.~(\ref{de:example}) of the following form: 
$
H(z;\varepsilon) = \sum_{j=0}^\infty h_j(z) \varepsilon^j \;.
$
\\
\noindent 
{\it Factorization.} 
Factorization of differential operators after $\varepsilon$-expansion means the following:
$$
\sum_{k=0}^p P_k(z;\varepsilon) \left( \frac{d}{dz} \right)^k
= 
\sum_{r=0} \Pi_{k=1}^{l_r \leq p } \left[ R_{k,r}(z) \frac{d}{dz}  + Q_{k,r}(z) \right] \varepsilon^{r} \;,
$$
where $ R_{k,r}(z)$ and $Q_{k,r}(z)$ are some rational functions. \\
{\it Linear parametrization.} 
Let us consider as illustration the  following differential equation  
\begin{eqnarray}
\left[ R_1(z) \frac{d}{dz}  + Q_1(z) \right] \left[ R_2(z) \frac{d}{dz}  + Q_2(z) \right] h(z) = F(z) \;.
\label{de}
\end{eqnarray}
Its iterated solution is:
\begin{equation}
f(z) = 
\int^{z} \frac{dt_3}{R_2(t_3)} \left[ \exp^{-\int_0^{t_3}\frac{Q_2(t_4)}{R_2(t_4)} dt_4} \right] 
\int^{t_3} \frac{dt_1}{R_1(t_1)} \left[ \exp^{-\int_0^{t_1}\frac{Q_1(t_2)}{R_1(t_2)} dt_2} \right] F(t_1) \;. 
\label{iterative:solution}
\end{equation}
In accordance with Eq.~(\ref{gmp}), this iterated integral can be written 
as multiple polylogarithm, if there is a new variable $\xi: \xi = \Psi(t)$, 
converting this  expression into ratio of polynomials \cite{KK2010}:  
\begin{eqnarray}
&& 
\int^z \frac{Q_i(t)}{R_i(t)} dt  =  \ln \frac{M_i(\xi)}{N_i(\xi)} \;, 
\quad
\left. \frac{dt}{R_2(t)} \right|_{t = t(\xi)} \frac{N_i(\xi)}{M_i(\xi)}  =  dx \frac{K_i(x)}{L_i(x)} \;, 
\label{abel}
\end{eqnarray}
where $M_i,N_i,K_i,L_i$ are polynomial functions. 
The existence of such a parametrization we called {\it linear parametrization}. 

{\it When does such parametrization exist?} 
To answer to this question, 
the non-homogeneous part of differential equation Eq.~(\ref{de:example})
should satisfy the system of linear PDE 
with {\it Factorization }
and 
{\it Linear parametrization}
in each order of $\varepsilon$:
\begin{eqnarray}
F(z;\varepsilon)  & = & \sum_{j=0}  f_j(z) \varepsilon^j\;,
\quad 
\Pi_{i=1}^r \left[ P_i(\xi) \frac{d}{d\xi}  + S_i(\xi) \right]  f_j(\xi)  = T_j(\xi)\;,
\end{eqnarray}
where $P_i, S_i, T$ are rational functions. 

\noindent 
{\bf Multivariable generalization} \\
Generalization of this technique for the Horn-type hypergeometric functions is straightforward: 
\begin{enumerate}
\item
Convert the system of linear PDE with polynomial coefficients into Pfaff form: 
\begin{eqnarray}
\sum_{J,k} P_{\vec{J};k}(\vec{a};\vec{z}) \frac{\partial}{\partial z_k} F(\vec{a};\vec{z}) = 0 
& \Rightarrow & 
\Biggl\{
d_k \omega_i(\vec{z}) = \Omega_{ij}^k(\vec{z}) \omega_j(\vec{z}) dz_k  \;,
\quad 
d_r \left[ d_k \omega_i(\vec{z}) \right] = 0 
\Biggr\} \;.
\nonumber 
\end{eqnarray}
\item
Find  the  values of parameters
when the last system of linear PDE can be converted into triangular form 
and when {\it Factorization} is valid.
\item
Find a linear parametrization: 
$
\mbox{validity of Eq.(\ref{abel}) for each variable} \;. 
$
\end{enumerate}

\noindent 
{\bf Simplification of the procedure of Factorization. }\\
To simplify the procedure of {\it Factorization} of differential operators, 
the following trick is very useful. 
Any Horn type hypergeometric function, defined by Eq.~(\ref{Phi}),
satisfies the system of linear PDE with polynomial coefficients: 
\begin{equation}
P_{\vec{L}}(\vec{z}) \frac{\partial^{\vec{L}}}{\partial \vec{z}} H(\vec{J};\vec{z}) = 0 \;, 
\label{homogeneous}
\end{equation}
where 
$ \frac{\partial^{\vec{L}}}{\partial \vec{z}} =  \frac{\partial^{l_1+\cdots+ l_k}}{\partial z_1^{l_1} \cdots \partial z_k^{l_k}}$
and 
$P_{\vec{L}}(\vec{z})$ are polynomial.
Moreover, there are linear differential operators that change the value of each parameter $J_a$ by $\pm 1$:
\begin{equation}
R_{a,\vec{K}}(\vec{z}) \frac{\partial^{\vec{K}}}{\partial \vec{z}} 
H(J_1, \cdots, J_{a-1}, J_a, J_{a+1}, \cdots, J_r;\vec{z})
= 
H(J_1, \cdots, J_{a-1}, J_a \pm 1, J_{a+1}, \cdots, J_r;\vec{z}) \;. 
\label{shift}
\end{equation}
In accordance with \cite{takayama}, the differential operators inverse to the operators defined
by Eq.~(\ref{shift}) can be constructed. 
Applying direct/inverse differential operators to the hypergeometric function 
the values of parameters can be changed by an arbitrary integer numbers:
\begin{equation}
Q_0(\vec{z}) H(\vec{J}+\vec{m};\vec{z}) =  \sum_{j=0}^r Q_{\vec{J}}(\vec{z}) \theta_{\vec{J}} H(\vec{J};\vec{z}) \;,
\end{equation}
where $\vec{m}$ is a  set of integers 
and 
$Q_0(\vec{z})$ and $Q_{\vec{J}}$ are polynomials and $r$ is the holonomic rank of system (\ref{homogeneous}). 
More details are given in \cite{MKL06}.

\noindent
{\bf Example II: $F_3$ hypergeometric function} \\
In \cite{f1:our,recent} we applied our algorithm for the construction 
of the $\varepsilon$-expansion of $F_1$ and $F_3$ Appell hypergeometric functions. 
The results of the $\varepsilon$ expansion for $F_1-$ \cite{f1:our} and $F_3-$ functions \cite{recent} 
around integer values of parameters are in agreement with results of \cite{xsummer}.
Let us consider the $\varepsilon$-expansion of the Appell hypergeometric function $F_3$
around rational values of parameters. 
The $\varepsilon$-expansion around this set of parameters does not follow from the algorithms described in \cite{MUW,W}. 

Let us consider the Appell hypergeometric function $F_3$:
\begin{eqnarray}
&& 
F_3\left(\frac{p_1}{q} \!+\! a_1 \varepsilon, \frac{p_2}{q}\!+\!a_2 \varepsilon, 
         \frac{r_1}{q}\!+\!b_1\varepsilon, \frac{r_2}{q}\!+\!b_2\varepsilon, 
          1\!-\!\frac{p}{q}+c\varepsilon; x,y
 \right)
\nonumber \\ && 
\qquad =
\sum_{m=0}^\infty
\sum_{n=0}^\infty
\frac{
\left( \frac{p_1}{q}\!+\!a_1\varepsilon \right)_{m} 
\left( \frac{p_2}{q}\!+\!a_2\varepsilon \right)_{n} 
\left( \frac{r_1}{q}\!+\!b_1\varepsilon \right)_{m} 
\left( \frac{r_2}{q}\!+\!b_2\varepsilon \right)_{n} 
}
{
\left( 1-\frac{p}{q}+c\varepsilon \right)_{m+n} 
}
\frac{x^m}{m!}
\frac{y^n}{n!} \;,
\end{eqnarray}
where $\{p_i,r_j,p,q \}$ are integers. 
Applying our technology step-by-step,
we find that the system of linear PDE
for the coefficient functions
is factorisable and has a triangular form only when
$p_j r_j = 0$ for $ j = 1,2.$
After that, the original system of linear PDE with polynomial coefficients 
is transformed 
into a linear system of PDE with algebraic coefficients. 
To convert this system into a class of linear PDE,
the linear parametrization should exist simultaneously for the each element of the singular locus of $F_3$: 
\begin{eqnarray}
\{ x \} 
\cup 
\{ 1-x \}
\cup
\{ y \} 
\cup
\{ 1-y \}
\cup
\{xy-x-y \}
\;,
\label{locus:f3}
\end{eqnarray}
as well as for the auxiliary functions $H_j, j=1,2,3$ defined as
\begin{eqnarray}
H_1(x) & = &  
(-1)^{\frac{s_1}{q}}
\left[ 
\frac{x^p}{(x-1)^{s_1+p}}
\right]^\frac{1}{q} 
\;, 
\quad 
H_2(x)  =   
(-1)^{\frac{s_2}{q}}
\left[ 
\frac{y^p}{(y-1)^{s_2+p}}
\right]^\frac{1}{q} 
\;, 
\nonumber \\
\qquad 
H_3(x,y) & = &  
(-1)^\frac{s_1+s_2}{q}
\left[ 
\frac{x^{s_2+p}y^{s_1+p}}
     {(xy-x-y)^{s_1+s_2+p}}
\right]^\frac{1}{q} 
\;, 
\label{h}
\end{eqnarray}
where 
$s_j = p_j + r_j$ and $j = 1,2$.
We find that the linear parametrization exists when: 
\begin{itemize}
\item
The functions $H_j$ are constant polynomial: $s_1=s_2=p=0$. It corresponds to \cite{MUW}.
\item
One of three functions $H_j$ are equal to $1$: $s_1=0, s_2=-p,$ and  $( 1 \leftrightarrow 2) .$ 
\item
Two of three  functions $H_j$ coincide: $s_1 \neq 0, s_2=p=0,$ and $( 1 \leftrightarrow 2) .$  
\end{itemize}
Unfortunately, for another physically interesting set of parameters \cite{KT}, 
we failed to rewrite the iterated integral in terms  of multiple polylogarithms.
For example, for $s_1=s_2=0$ and $p \neq 0$, the statement about the existence of a linear parametrization is equivalent to 
the existence of three different (rational) polynomial functions of two variables $P_1(x,y),P_2(x,y)$ and $P_3(x,y)$, such that
\begin{equation}
P_1^q + P_2^q + P_3^q = 1\;,
\label{P1P2P3}
\end{equation}
where $q$ is integer and $q \geq 2$.
To our knowledge, this equation has a solution only in the class of elliptic functions. 
However, the finite part of the $F_3$-function with this set of parameters is expressible in terms of polylogarithms \cite{Davydychev:box}.

As result we got, that only for the two cases:
\begin{eqnarray}
&&
F_3\left(I_1 + \frac{p_1}{q} + a_1 \varepsilon, I_2+a_2 \varepsilon, I_3+b_1\varepsilon, I_4+b_2\varepsilon,
        I_5+\frac{p_1}{q}+c\varepsilon; x,y \right) \;,
\label{result1}
\\ &&
F_3\left(I_1 + \frac{p_1}{q} + a_1 \varepsilon, I_2+a_2 \varepsilon, I_3+b_1\varepsilon, I_4+b_2\varepsilon,
        I_5+c\varepsilon; x,y \right) \;,
\label{result2}
\end{eqnarray}
where $I_j, p_1, q$ are integers, 
the analytical coefficients of the $\varepsilon$-expansion of $F_3$ hypergeometric function are explicitly expressible 
in terms of multiple polylogarithms \cite{2dm}.

\noindent 
{\bf Hypergeometric Functions vs. Feynman Diagram} \\
These two building blocks, {\it Factorization}
and {\it Linear parametrization},
are sufficient to rewrite an iterative solution of system of linear PDE in terms of multiple polylogarithms.
It is a  cornerstone of all modern multiloop analytical evaluations of 
master-integrals in QCD 
and our results are in full agreement with available QCD calculations \cite{DE:1,DE:2,DE:3}. 


\noindent 
{\bf Conclusion}: \\
The algorithm described in \cite{our,KK2010,recent} 
has been applied to the construction of  the analytical coefficients of $\varepsilon$-expansion 
of Horn-type hypergeometric functions of two variables \cite{appell}
as well as Mellin-Barnes integrals \cite{MB}.
In particular, we analyzed the  following linear system:
\begin{eqnarray}
U_{0} \theta_{11} 
\omega(\vec{z};\varepsilon) 
& = &
\Biggl\{
U_{1} \theta_{12}
\!+\!
P_1 \theta_1
\!+\!
P_2 \theta_2
\!+\!
P_0
\Biggr\}
\omega(\vec{z};\varepsilon) 
\;,
\nonumber \\
\quad 
T_{0} \theta_{22} 
\omega(\vec{z};\varepsilon) 
& = & 
\Biggl\{
T_{1} \theta_{12}
\!+\!
R_1 \theta_1
\!+\!
R_2 \theta_2
\!+\!
R_3 
\Biggr\}
\omega(\vec{z};\varepsilon) 
\;,
\label{system}
\end{eqnarray}
where 
$\vec{z}  =  (z_1,z_2)$ are independent variables, 
$
\theta_j = z_j \partial_{z_j} \;,   j=1,2 \;,  
$
and
$
\theta_{i_1\cdots i_k} = \theta_{i_i} \cdots \theta_{i_k}.
$
The functions  
$G_0 \equiv \{U_0,T_0,U_{1},T_{1} \}$ are polynomial of variables $z_1$ and $z_2$:
\begin{eqnarray}
G_0 = \sum_{i,j=0} \sigma_{i,j} z_1^i z_2 ^j \;, 
\end{eqnarray}
all other function $E_0 \equiv \{P_i, R_j\}$ are polynomial of variables $z_1,z_2$ and $\varepsilon$:
\begin{eqnarray}
E_0 = \sum_{i,j,k=0} \gamma_{i,j;k} z_1^i z_2 ^j \varepsilon^k \;, 
\end{eqnarray}
and 
$\sigma_{i,j}$ 
and 
$\gamma_{i,j;k}$ 
are rational. 
The analytical structure of the coefficients $h^{(r)}_j(\vec{z})$ defined via 
Laurent expansion around $\varepsilon=0$ of the solution 
$\omega^{(r)}(\vec{z};\varepsilon)$
of this system have been analyzed
\begin{eqnarray}
\omega^{(r)}(\vec{z};\varepsilon)  = \sum_{j} h^{(r)}_j(\vec{z}) \varepsilon^j \;. 
\end{eqnarray}
for the physically interesting set of parameters \cite{KT}. 
In particular, the hypergeometric functions considered in \cite{MUW,W} 
correspond to a system of linear PDE (\ref{system})
with polynomial coefficients and with singularity locus
\begin{eqnarray}
L:=
\{ z_1 \} 
\cup 
\{ U_0 \}
\cup
\{ z_2 \} 
\cup
\{ T_0 \}
\cup 
\{U_0 T_0 - U_1 T_1 \}
\;,
\label{locus}
\end{eqnarray}
where 
\begin{eqnarray}
U_i(\vec{z};\vec{a}) & = & a_{0,i} + a_{1,i} z_1 + a_{2,i} z_2 \;, 
\quad 
T_j(\vec{z};\vec{a})  =  b_{1,j} z_1 + b_{2,j} z_2 + b_{3,j} z_1 z_2 \;, 
\end{eqnarray}
$i,j=1,2$ and 
$a_{k,j},b_{k,i} \in \{ 0, \pm 1\}$.
See also \cite{simplify,bogner}.

The $\varepsilon$-expansion around rational values of parameters
with one unbalanced rational parameter, 
corresponds to a system of linear PDE with algebraic or elliptic coefficients. 
Imposing only {\it Factorization} conditions gives rise to 
iterative integrals with algebraic functions, that in general,
are not expressible in terms of {\it multiple polylogarithms}. 
Only when additional {\it Linear parametrization} conditions are valid, 
we are able to rewrite the results of the integration in terms of 2-dimensional polylogarithms \cite{2dm}. 
The {\it Linear parametrization} should  exist simultaneously 
for the each element of singular locus, Eq.~(\ref{locus}), 
of the differential system Eq.~(\ref{system})
and for algebraic functions defined as q-roots of ratios of elements of $L$:
$\left(L_i/(1-L_i) \right)^\frac{1}{q}$ and/or
$\left((L_i L_j) /(L_i + L_j - L_i L_j) \right)^\frac{1}{q}$ 
 (see Eq.~(\ref{h})). 
It is in agreement with the one-variable case analysed in \cite{KK2010}. 

We got, see also \cite{KK2010}, that even when the finite part of a hypergeometric function is expressible in terms 
of multiple polylogarithms (existence of Liouvillian solution of a linear system of PDE in $d=4$) 
it does not follow that higher order terms of the $\varepsilon$-expansion are expressible 
in terms of multiple polylogarithms, too. 

\noindent
{\bf Acknowledgments} \\
MYK would like to thank the organizers of the conference 
{\it Loops and Legs in Quantum Field Theory} 
for the invitation and for creating such a stimulating atmosphere.
Our thanks to all participants, 
but especially to 
J.~Bl\"umlein, 
D.~Broadhurst, 
A.~Grozin,
M.~Hoffman, 
D.~Kreimer, 
S.~Moch 
and 
S.~Weinzierl
for useful discussion.
MYK is grateful to J.~Gluza for assistance with \cite{ambre}
and to P.~Bolzoni and I.~Bierenbaum for carefully reading this manuscript.

\end{document}